\documentclass[preprint,showpacs,preprintnumbers,amsmath,amssymb]{revtex4}

\usepackage{graphicx}
\usepackage{dcolumn}
\usepackage{bm}

\begin{document}

\preprint{APS/123-QED}

\title{Measurement of $K^+ \rightarrow \pi^0 \pi^0 e^+ \nu$ ($K_{e4}^{00}$) 
decay using stopped positive kaons}

\author{
S.~Shimizu$^{1}$, K.~Horie$^{1}$\cite{Pek}, M.~Aliev$^{2}$, Y.~Asano$^{3}$,
T.~Baker$^{4}$, P.~Depommier$^{5}$, M.~Hasinoff$^{6}$,
Y.~Igarashi$^{4}$, J.~Imazato$^{4}$, A.P.~Ivashkin$^{2}$,
M.M.~Khabibullin$^{2}$, A.N.~Khotjantsev$^{2}$, Y.G.~Kudenko$^{2}$,
A.~Levchenko$^{2}$, G.Y.~Lim$^{4}$, J.A.~Macdonald$^{7}$\cite{jam},
O.V.~Mineev$^{2}$, C.~Rangacharyulu$^{8}$ and S.~Sawada$^{4}$~~\\ 
(KEK-E470 Collaboration)\\
}
\affiliation{
$^{1}$Department of Physics, Osaka University, Osaka 560-0043, Japan\\ 
$^{2}$Institute for Nuclear Research, Russian Academy of Sciences, Moscow 117312, Russia \\
$^{3}$Institute of Applied Physics, University of Tsukuba, Ibaraki 305-0006, Japan \\ 
$^{4}$IPNS, High Energy Accelerator Research Organization (KEK), Ibaraki 305-0801, Japan \\ 
$^{5}$Laboratoire de Physique Nucl\'eaire, Universit\'e de Montr\'eal,
Montreal, Qu\'ebec, Canada H3C 3J7 \\
$^{6}$Department of Physics and Astronomy, University of British Columbia, Vancouver, Canada V6T 1Z1 \\
$^{7}$TRIUMF, Vancouver, British Columbia, Canada V6T 2A3 \\
$^{8}$Department of Physics, University of Saskatchewan, Saskatoon, Canada S7N 5E2 \\
}%

\date{\today}

\begin{abstract}
The $K^+ \rightarrow \pi^0 \pi^0 e^+ \nu$ ($K_{e4}^{00}$) decay has been
measured with stopped positive kaons for a data sample of
216 events. A comparison of the observed spectra
with a Monte Carlo simulation determined the $K_{e4}^{00}$ form factor.
The results are compatible with the $K^+ \rightarrow \pi^+ \pi^- e^+ 
\nu$ data, as estimated from the $\Delta I=1/2$ rule. We also established 
that the $K_{e4}^{00}$ channel can be used to determine the $\pi$-$\pi$ 
scattering lengths.
\end{abstract}

\maketitle
$K\rightarrow\pi \pi e \nu$ ($K_{e4}$) decays are the subject of 
continuing interest for a variety of reasons~\cite{cab65,pai68,cal66}. 
The decay form factors 
serve as constraints on the parameters of the Lagrangians in the framework 
of Chiral Perturbation Theory (ChPT). More importantly, they are
clean sources of $\pi$-$\pi$ pairs at low energy, from which one can 
deduce the scattering lengths of significance for models of hadron 
dynamics.

Recently, the BNL-E865 group reported on the precise measurement of the $K^+ 
\rightarrow \pi^+ \pi^- e^+ \nu$ ($K_{e4}^{+-}$) decay, yielding the decay 
form factors as well as the $S$-wave, isospin zero $\pi$-$\pi$ 
scattering length $a_0^0$~\cite{pis01}. 
In addition to the $K_{e4}^{+-}$ decay, there are two other $K_{e4}$ decays,
$K^+ \rightarrow \pi^0 \pi^0 e^+ \nu$ 
($K_{e4}^{00}$) and $K_L \rightarrow \pi^- \pi^0 e^+ \nu$ 
($K_{e4}^{-0}$)~\cite{mak93}. Since these $K_{e4}$ decays are theoretically 
related to each other by simple isospin arguments~\cite{ber67}, 
comprehensive experimental studies for 
all channels provide deep insight into the understanding of $K_{e4}$ 
physics. Among the three $K_{e4}$ 
decays, the $K_{e4}^{00}$ channel is the simplest because the decay 
kinematics can be described by only one form factor in view of the 
identity of the two $\pi^0$s in the final state. Thus far, only 37 
$K_{e4}^{00}$ events have been observed using an in-flight-$K^+$ 
technique~\cite{bol86,bar88}.  A more accurate 
measurement of the $K_{e4}^{00}$ channel was desirable as an additional 
independent check on the $\pi$-$\pi$ scattering length and the $\Delta 
I=1/2$ rule. 

In this paper, we present a new measurement of $K_{e4}^{00}$ decay.
The experiment used a stopped $K^+$ beam in conjunction with a 12-sector
iron-core superconducting spectrometer. 
We have been able to make
a first attempt to deduce the $q^2$ dependent terms of the form factor and 
determine the $\pi$-$\pi$ scattering length in a nearly model independent 
manner.

The experiment was performed at the KEK 12 GeV proton synchrotron.
The experimental apparatus was based on the E246 experiment, a search for 
the $T$-violation in $K^+ \rightarrow \pi^0 \mu^+ \nu$ decay 
($K_{\mu3}$)~\cite{main:99,nim}.  Besides the $T$-violation search, 
spectroscopic studies for various decay modes have been successfully 
performed~\cite{shi00}. After the E246 data 
collection was completed in 2000, the E470 experiment to measure direct 
photon emission in the 
$K^+\rightarrow \pi^+ \pi^0 \gamma$ ($K_{\pi2 \gamma}$) decay was 
carried out~\cite{ali03}, and the $K_{e4}^{00}$ events were 
simultaneously recorded.

A separated 660 MeV/$c$ $K^+$ beam was stopped in an active target system
located at the center of the spectrometer. The $K_{e4}^{00}$ events
were identified by analyzing the $e^+$ momentum with the spectrometer
and detecting the four photons in the CsI(Tl) calorimeter. Charged
particles from the target were momentum-analyzed 
using multiwire proportional chambers, as well as an array of ring
scintillators. The $e^+$s were separated from $\mu^+$s by determining
the mass squared ($M_{\rm TOF}^2$) of the charged particles from the
measured time-of-flight between TOF1 and TOF2 counters. Counter TOF1
surrounded the active target and counter TOF2 was located at the exit
of the spectrometer. The photon energy and hit position were obtained,
respectively, by summing the energy deposits and taking the
energy-weighted centroid of the crystals sharing the shower. 

$K_{e4}^{00}$ decays at rest were extracted by the following procedure.
The $K^+$ decay time, defined as the $e^+$ signal in the TOF1 counter, was
required to be more than 1.4 ns later than the $K^+$ arrival time as measured
by the ${\rm \check{C}}$erenkov counter. This reduced the fraction of $K^+$
decay in-flight contamination to the level of $10^{-3}$ of the
stopped $K^+$s. Events with $e^+$ scattering on the magnet pole faces
were eliminated by requiring the hit position in the ring counters to be
consistent with the charged particle track. The selection of positrons
required $-5000<M_{\rm TOF}^2<5000$ MeV$^2/c^4$. Events with four
photon clusters coming from two $\pi^0$s were selected, while events
with other photon cluster number were rejected. Photon
conversion backgrounds in the active target system could be removed by
selecting only events in which there was one charged particle hit in
TOF1 as well as in the ring counters. 

Since there are three possible combinations to form two $\pi^0$s from 
the four photons, the $K_{e4}^{00}$ events were reconstructed by 
introducing a quantity $Q^2$,
\begin{eqnarray}
        Q^2=(M_{\pi^0_1}-M_{1})^2/\sigma_{M_1}^2&+&
        (M_{\pi^0_2}-M_{2})^2/\sigma_{M_2}^2 \nonumber \\
         &+&({\rm cos}\theta_{\pi^0 \pi^0}^{\rm meas}-
        {\rm cos}\theta_{\pi^0 \pi^0}^{\rm calc} -\delta)^2/\sigma_{\delta}^
2, \label{q2}
\end{eqnarray}
where $M_{\pi^0}$ is the invariant mass of the selected pair (the
first $\pi^0$ has higher energy) and 
$\theta_{\pi^0 \pi^0}$ is the opening angle between the $\pi^0$s.
The superscripts MEAS and CALC stand for the measured angle and the 
angle calculated from the measured photon energies, respectively. The 
photon pair
with the minimum, $Q^2_{\rm min}$, among the three combinations
was adopted as the correct pairing. The $\sigma$ and offset values 
in each term are $\sigma_{M_1}=12.74$ MeV/$c^2$, $\sigma_{M_2}
=15.42$ MeV/$c^2$, $M_1=124.5$ MeV/$c^2$, $M_2=113.5$ MeV/$c^2$,
$\delta=-0.019$, and $\sigma_{\delta}=0.336$. The choice of Eq.~(\ref{q2}) 
and these parameters were determined to maximize the probability for
the correct pairing using a GEANT based Monte Carlo code. Because of
the shower leakage from the calorimeter holes, the $\sigma$ and offset
values differ from the ideal ones. The correct pairing probability was
estimated to be 96\% from the simulation. 

\begin{figure}
\includegraphics{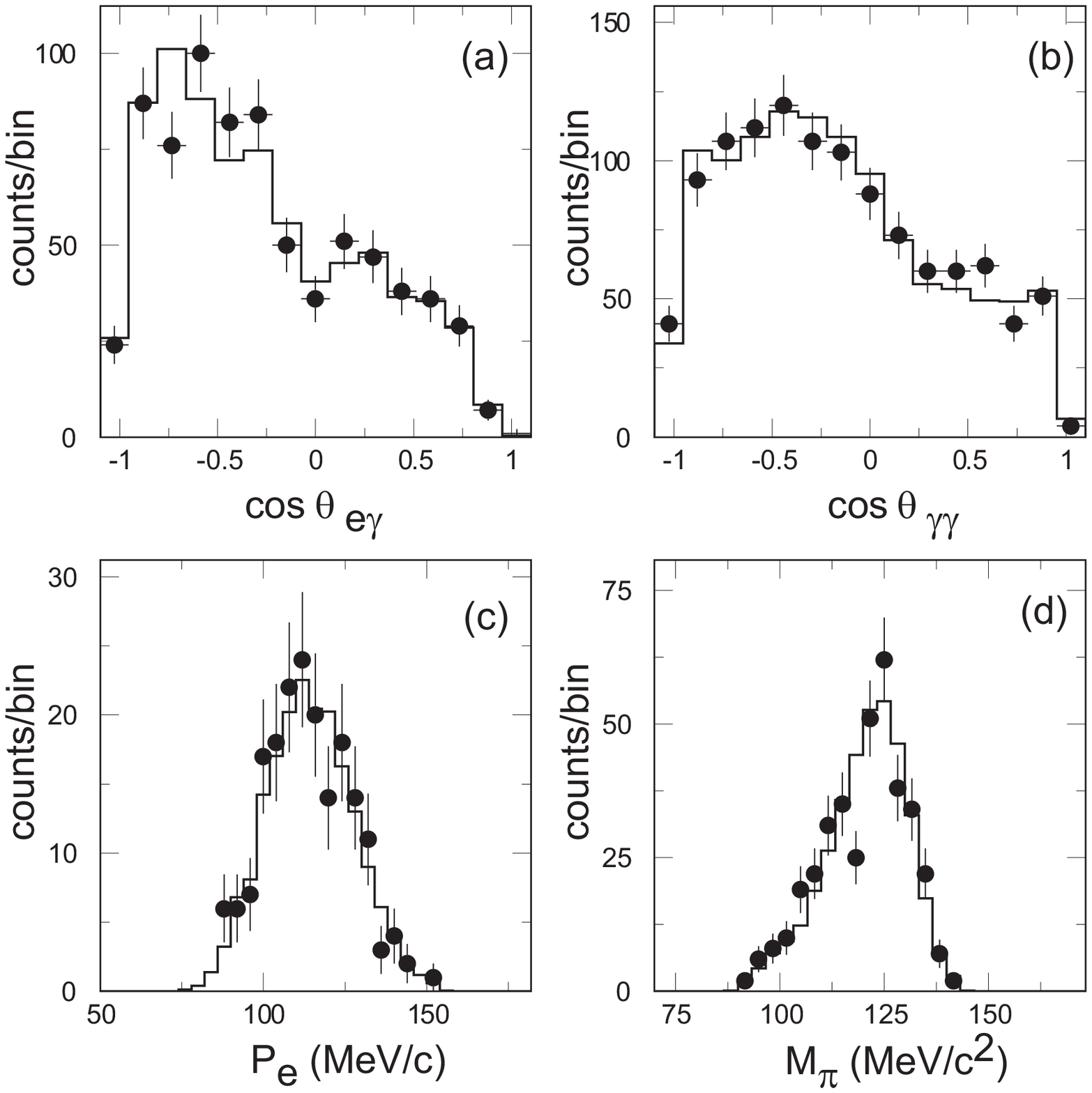}
\caption{\label{fg:brem} $K_{e4}^{00}$ spectra: (a) cos$\theta_{e^+
\gamma}$, (b) cos$\theta_{\gamma \gamma}$, (c) $e^+$ momentum,
(d) $M_{\pi^0}$ for the experimental data (dots) and a Monte Carlo
simulation of $K_{e4}^{00}$ (histogram). Four and six combinations for
cos$\theta_{e^+ \gamma}$  and cos$\theta_{\gamma \gamma}$,
respectively, were accumulated in the same figures. Both $M_{\pi^0_1}$
and $M_{\pi^0_2}$ values were also accumulated in (d).   
}
\end{figure}

Since backgrounds due to $K^+ \rightarrow 
\pi^0 e^+ \nu$ ($K_{e3}$) and $K^+ \rightarrow \pi^0 e^+ \nu \gamma$
($K_{e3 \gamma}$) decays with accidental photons, photons split
into multiple clusters, and bremsstrahlung photons do not satisfy the
$K_{e4}^{00}$ kinematics, the cut of $Q_{\rm min}^2<3$ essentially
removed these backgrounds.  The fraction of the accidental
background in the calorimeter was estimated to be 1.0\% by changing the
CsI(Tl) TDC gate widths. Using the Monte Carlo simulation, background
fractions in the $K_{e4}$ sample due to the bremsstrahlung photons and
photons split into multiple clusters were obtained to be 
0.02\% and 0.06\%, respectively.  The total level of the backgrounds was
thus derived to be 1.1\% which was dominated by $K_{e3 \gamma}$ with
accidental photons. The opening angle distributions between 
$e^+$ and four $\gamma$s ($\theta_{e^+\gamma}$) and between each
photon ($\theta{\gamma \gamma}$) for the selected $K_{e4}$ events were
compared to the simulation to confirm the correctness of the
simulation conditions, as shown in Fig.~\ref{fg:brem}(a)(b). No
deviation of $\theta_{e^+ \gamma}$ and $\theta_{\gamma \gamma}$ from
the simulation at small angles was observed, supporting the correct
estimation of these background fractions. After these selection
conditions, the number of good $K_{e4}^{00}$ events was found to be
216. The $e^+$ momentum and $\pi^0$ invariant mass spectra of the
selected $K_{e4}^{00}$ events are shown in
Fig.~\ref{fg:brem}(c)(d). 

\begin{figure}
\includegraphics{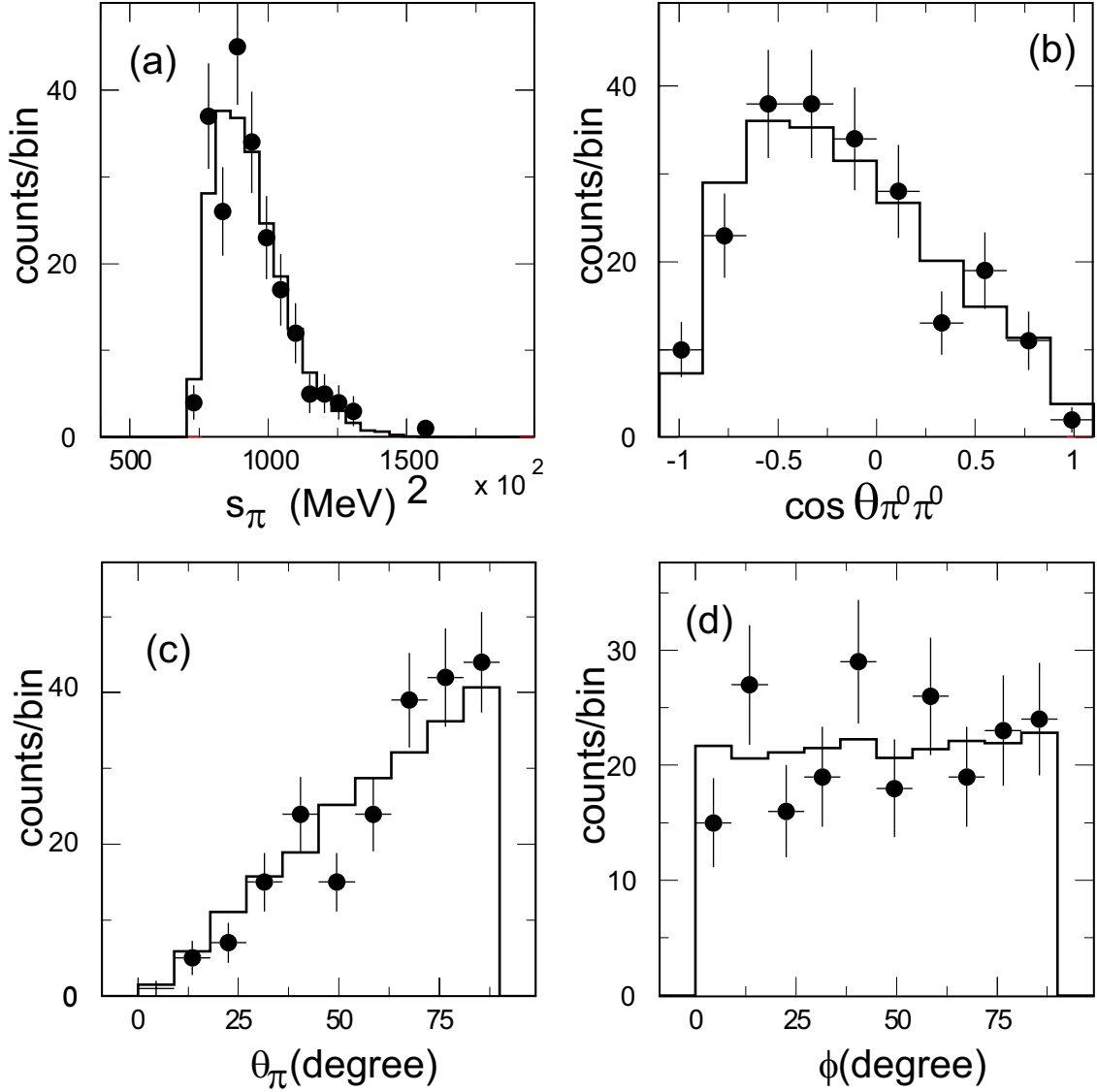}
\caption{ \label{fg:ke4} $K_{e4}^{00}$ spectra: (a) $s_{\pi}$,
(b) cos$\theta_{\pi^0 \pi^0}$, (c) $\theta_{\pi}$, (d) $\phi$, for the
experimental data (dots) and the Monte Carlo simulation of
$K_{e4}^{00}$ with the best fitted $f'/f_0$ and $f''/f_0$ parameters
(histograms).  
}
\end{figure}

\newcounter{aa}
In general, the $K_{e4}$ kinematics can be written as a set of five 
configuration variables~\cite{cab65,cal66}: \setcounter{aa}{1}(\Roman{aa})
$s_{\pi}=M_{\pi \pi}^2$, the effective mass  squared of the dipion
system, \setcounter{aa}{2} (\Roman{aa}) $s_l=M_{e \nu}^2$, the
effective mass squared of the dilepton system,
\setcounter{aa}{3}(\Roman{aa}) $\theta_{\pi}$, the angle made by the
$\pi^0$ in the dipion center of mass with respect to the kaon
direction, \setcounter{aa}{4}(\Roman{aa}) 
$\theta_{l}$, the angle made by the $e^+$ in the dilepton center of
mass with respect to the kaon direction, and
\setcounter{aa}{5}(\Roman{aa}) $\phi$, the angle between the decay
planes formed by the dipion and dilepton systems. Because of the
identity of the two $\pi^0$s, $\theta_{\pi}$ and $\phi$ can be defined
in the region of $0<\theta_{\pi}<\pi/2$ and $0<\phi<\pi/2$. The most
general form of the $K_{e4}$ matrix element in terms of the  hadronic
vector and axial vector current contributions $V^{\mu}$ and $A^{\mu}$
is given by~\cite{cab65}  
\begin{eqnarray}
        M&=&\frac{G_F}{\sqrt{2}}V_{us} \bar{u}(p_{\nu})\gamma_{\mu}(1-\gamma_5)
        v(p_e)(V^{\mu}-A^{\mu}), \\
        A^{\mu}&=&FP^{\mu}+GQ^{\mu}+RL^{\mu}, \\
        V^{\mu}&=&H\epsilon^{\mu\nu\rho\sigma}L_{\nu}P_{\rho}Q_{\sigma},
\end{eqnarray}
where $P=p_1+p_2$, $Q=p_1-p_2$, $L=p_e+p_{\nu}$, and $p_1$, $p_2$, $p_e$,
$p_{\nu}$ are the four-momenta of $\pi_1$, $\pi_2$, $e^+$, $\nu$ in units
of $m_K$, respectively. The form factors $F$, $G$, $R$, and $H$ are
dimensionless functions of $s_{\pi}$. The $R$ term enters the decay
distribution multiplied by $m_e^2$ and is therefore negligible.
In general, $S$ and $P$ waves in the dipion system were considered
to contribute to the $K_{e4}$ decays~\cite{pis01}. However, odd angular 
momenta $\pi$-$\pi$ states cannot be present in the $K_{e4}^{00}$ 
channel due to symmetry arguments. Therefore, only a contribution 
from the $S$-wave component is possible and
the decay kinematics are described only by the $F$ form factor~\cite{ber67}.
Here, a parameterization~\cite{amo99} 
\begin{eqnarray}
	F=f_0+f'q^2+f''q^4,  \label{eq:f} 
\end{eqnarray}
is employed, where $q^2$ is the momentum transfer squared $q^2=s_\pi/(
4m^2_{\pi})-1$ to the lepton current. The observed $K_{e4}^{00}$
spectra ($s_{\pi}$, cos$\theta_{\pi^0 \pi^0}$, $\theta_{\pi}$, $\phi$)
which are sensitive to the $K_{e4}^{00}$ form factor are shown in
Fig.~\ref{fg:ke4}. 

The $f'/f_0$ and $f''/f_0$ values could be extracted by 
fitting the observed spectra to the Monte Carlo simulation. 
The simulation data were generated 
according to Pais and Treiman~\cite{pai68} and analyzed in the same manner 
as the experimental sample. The simulation spectra were then obtained as a 
function of the form factor by re-weighting 
the reconstructed simulation events using the generated Monte Carlo 
kinematic variables. A program based on MINUIT minimized $\chi^2$ and 
estimated the error of the parameters. The reproducibility of the 
experimental conditions in the simulation was carefully checked using 
$K_{\pi 2}$ and $K_{\pi 3}$ decays.

In the case of the $K_{e4}^{+-}$ decay, the $\pi$-$\pi$ scattering length
can be determined from the phase-shift difference between the $S$ and 
$P$ waves~\cite{pis01}. However, a similar procedure would not be of
use for the $K_{e4}^{00}$ decay.
Instead, one can assume that the $s_{\pi}$ 
dependence of $F$ is described by an $S$-wave phase-shift parameter 
$\delta_0^0$~\cite{cab65}:
\begin{eqnarray}
	F\propto (1/\beta){\rm sin}[\delta_0^0(s_{\pi})]
	e^{i\delta_0^0(s_\pi)}, \label{eq:cab}
\end{eqnarray}
where $\beta=(1-4m_{\pi}^2/s_{\pi})^{1/2}$ and the $s_{\pi}$ dependence of
the phase shift is given by the Chew-Mandelstam 
effective-range formula~\cite{che60}:
\begin{eqnarray}
	{\rm cot}\delta_0^0=1/(\beta a_0^0)+(2/\pi){\rm ln}
	[\sqrt{s_{\pi}}/(2m_{\pi})(1+\beta)].
\end{eqnarray}
To determine the $a_0^0$ value, this $F$ form factor was used to generate 
the $K_{e4}^{00}$ events in the simulation. Since the $f'/f_0$ and $f''/f_0$ 
parameters are numerically determined from $a_0^0$ by fitting 
Eq.~(\ref{eq:cab}) to Eq.~(\ref{eq:f}), it is possible to
check the consistency of the theories by comparing the results of the 
($f'/f_0$, $f''/f_0$) and $a_0^0$ analyses.

\begin{figure}
\includegraphics{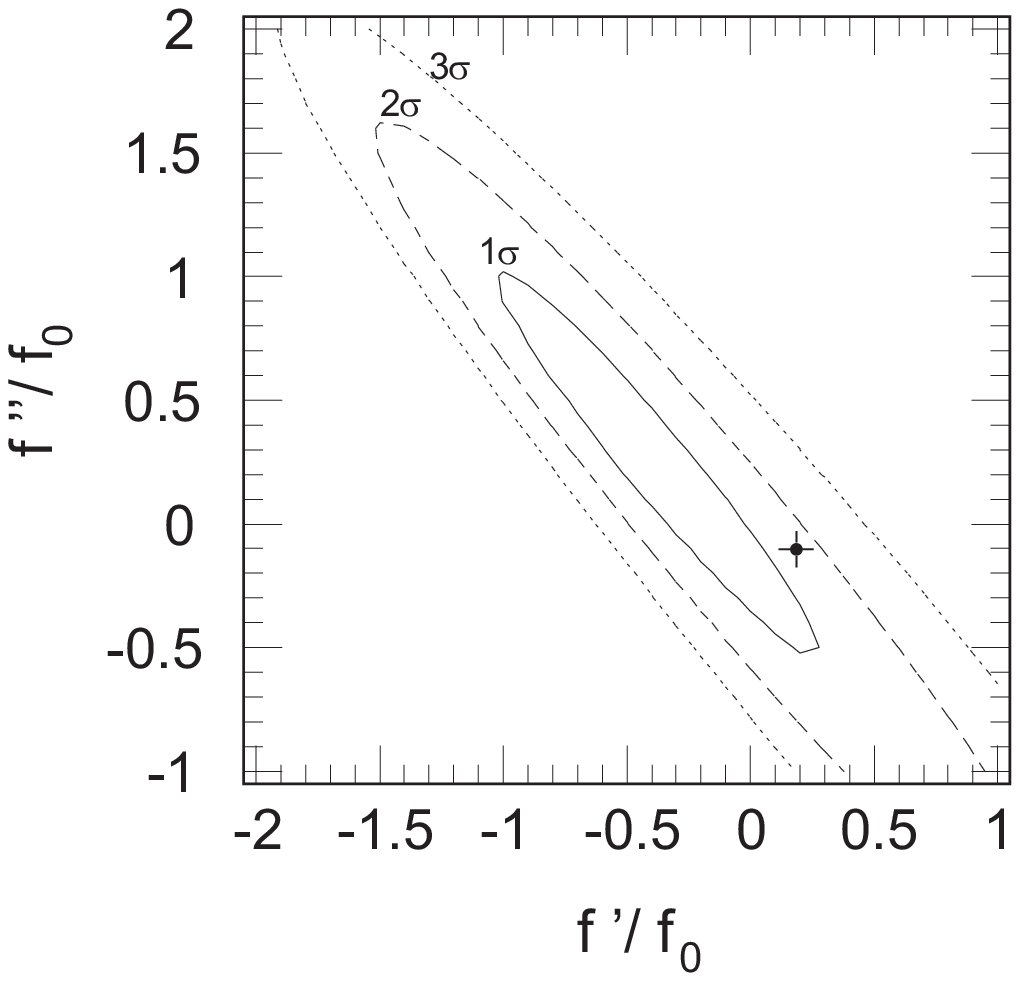}
\caption{\label{fg:chi2} $\chi^2$ contour plot in the ($f'/f_0,f''/f_0$) space.
Contour levels of 1-, 2-, and 3$\sigma$ constraints are drawn.
The $K_{e4}^{+-}$ experimental result by the BNL-E865 group is 
also plotted.
  }
\end{figure}

The $f'/f_0$ and $f''/f_0$ values were determined from the density
distribution of the two 
observables, $s_{\pi}$ and $\theta_{\pi^0 \pi^0}$.
The experimental distribution $\rho(s_{\pi}, \theta_{\pi^0
\pi^0})$ was fitted to a simulation spectrum with 
$f'/f_0$ and $f''/f_0$ being free parameters. They were obtained to be
$f'/f_0=-0.45^{+0.75}_{-0.59}$ and $f''/f_0=0.32^{+0.72}_{-0.89}$. The
reduced $\chi^2$ was 0.91. Fig.~\ref{fg:chi2} shows the $\chi^2$ contour 
plot in the ($f'/f_0$, $f''/f_0$) space. The histograms in  
Fig.~\ref{fg:ke4} shows the simulation spectra with the best fit
parameters, which are in good agreement with the experimental data.  
Using these form factors for the acceptance calculation, we obtained
$\Gamma(K_{e4}^{00})= [(3.56\pm0.26(stat.)^{+7.51}_{-2.92}(syst.)]\times
10^3$ $s^{-1}$ by normalizing it to the theoretical value of internal
bremsstrahlung in the $K_{\pi2 \gamma}$ decay~\cite{daf92}. The large
systematic error from the uncertainty of the form factors (i.e.,
uncertainty of the $e^+$ momentum distribution) was due to the fact
that our experiment covered a very narrow region of phase space and
the error of the detector acceptance became large. As shown in
Fig.~\ref{fg:ke4}, the experimental $\theta_{\pi}$ and $\phi$
distributions were well accounted for by the simulations without any
$D$-wave component. Thus, the data are compatible with an $S$-wave
only description. The reduced $\chi^2$ was 0.92 for $\theta_{\pi}$ and
1.19 for $\phi$. This feature also supports the previous $K_{e4}^{+-}$
measurement in which the $D$-wave contribution was assumed to be
negligible~\cite{pis01}.  

Then, using Eq.~(\ref{eq:cab}) as the $F$ form factor, the $a_0^0$ 
value which gave the best fit to the experimental $\rho({s_{\pi}, 
{\rm cos}\theta_{\pi^0 \pi^0}})$ distribution was determined to be  
$a_0^0=0.45\pm0.43$ in units of $m_{\pi}$. The reduced $\chi^2$ was
0.91. While the large error precludes a meaningful comparison with 
results from other channels, 
a well designed experiment at the future high intensity facilities
such as J-PARC will be able to make a significant contribution to the
determination of the $a_0^0$ parameter.
The $f'/f_0$ and $f''/f_0$ parameters were determined from this
$a_0^0$ value to be $f'/f_0=-0.30^{+0.29}_{-0.33}$ and
$f''/f_0=0.12^{+0.18}_{-0.12}$, which are in agreement with the
($f'/f_0$, $f''/f_0$) fitting.

The $\Delta I=$1/2 rule predicts that the $q^2$ dependence of $K_{e4}^{00}$
and $K_{e4}^{+-}$ form factors are identical. As seen in Fig.~\ref{fg:chi2}, 
our $K_{e4}^{00}$ result overlaps with the $K_{e4}^{+-}$ form factor 
reported by the BNL-E865 group within two standard
deviations~\cite{pis01}. The $f'/f_0$ values were obtained to be
$-0.19^{+0.19}_{-0.17}$ and $-0.11^{+0.19}_{-0.17}$ for $f''/f_0=0$
and $-0.10$ ($K_{e4}^{+-}$ result), respectively, which also overlaps
with the $K_{e4}^{+-}$ result within the error~\cite{pis01}. The
$\Delta I=$1/2 rule also predicts a simple relation for the decay widths
$\Gamma(K_{e4}^{00})=1/2\Gamma(K_{e4}^{+-})-1/4\Gamma (K_{e4}^{-0}) 
$~\cite{ber67}. The world averages for the $K_{e4}^{+-}$ and  
$K_{e4}^{-0}$ channels~\cite{pdb02} give $\Gamma(K_{e4}^{00})=(1.40\pm0.04)
\times 10^3$ $s^{-1}$. Using the $K_{e4}^{+-}$ form factor~\cite{pis01} 
for the acceptance calculation of the present experiment to reduce the
systematic error, we deduced  $\Gamma(K_{e4}^{00})= [1.85\pm
0.13(stat.)\pm 0.24(syst.)] \times 10^3$ $s^{-1}$. The average of the
previous $K_{e4}^{00}$ experiments, under the assumption of constant
form factor, is quoted as $\Gamma(K_{e4}^{00})=(1.70 \pm 0.32)\times
10^3$ $s^{-1}$~\cite{pdb02}. While these numbers are all consistent with
each other, the large systematic error of the $\Gamma(K_{e4}^{00})$
value determined in the present experiment
precludes a meaningful comparison with other results.
Further measurements will be essential before we can draw firm conclusions 
on the $\Delta I=3/2$ contributions.

In conclusion, we have performed a new measurement of one of the
$K_{e4}$ channels $K^+ \rightarrow \pi^0 \pi^0 e^+ \nu$
($K_{e4}^{00}$) using stopped positive kaons. 
The data sample of 216 events is, while a factor of six improvement
over the world data set, still very limited.
The detector response
and acceptance functions were evaluated by the  Monte Carlo
simulation. The backgrounds due to $K_{e3}$ and $K_{e3 \gamma}$ decays
were estimated to be 1.1\%. The $F$ form factor of the
$K_{e4}^{00}$ decay was determined by fitting the observed
$\rho(s_{\pi}, \theta_{\pi^0 \pi^0}$) distribution to the simulation. 
A first attempt to deduce the $S$-wave $\pi$-$\pi$ scattering length 
from the $\rho(s_{\pi}, \theta_{\pi^0 \pi^0}$) distribution was 
made. The good reproducibility of the experimental $\theta_{\pi}$ and
$\phi$ spectra by the simulation indicates that the data are
compatible with an $S$-wave only description. These results are
consistent with the previous $K^+ \rightarrow \pi^+ \pi^- e^+ \nu$
measurement within errors. 
By this measurement, despite the limited statistics, we established the
physics potential of the $K_{e4}^{00}$ channel.

This work has been supported in Japan by a Grant-in-Aid from the Ministry 
of Education, Culture, Sports, Science and Technology, and by JSPS; in 
Russia by the Ministry of Science and Technology, and by the Russian 
Foundation for Basic Research.

\newpage


\newpage

\newpage

\newpage

\begin{references}
\bibitem[$\ast$]{Pek} Present address: IPNS, High Energy Accelerator Research Organization (KEK), Ibaraki 305-0801, Japan
\bibitem[$\dag$]{jam} Deceased 22 May 2003.
\bibitem{cab65} N.~Cabibbo and A.~Maksymowicz, Phys. Rev. {\bf 137} B438
        (1965);~Phys. Rev. {\bf 168} 1926 (1968). 
\bibitem{pai68} A.~Pais and S.~B.~Treiman, Phys. Rev. {\bf 168} 1858 (1968).
\bibitem{cal66} C.~Callan and S.~Treiman, Phys. Rev. Lett. 
	{\bf 16} 153 (1966);~S.~Weinberg, Phys. Rev. Lett. {\bf 17} 
	336 (1966);~G.~Amoros, J.~Bijnens, and P.~Talavera, Nucl. Phys.
	{\bf B585} 293 (2000);~G.~Colangelo, J.~Gasser, and H.~Leutwyler, 
	Phys. Rev. Lett. {\bf 86} 5008 (2001).  
\bibitem{pis01} S.~Pislak {\em et al.}, Phys. Rev. Lett. 
	{\bf 87} 221801 (2001).  
\bibitem{mak93} G.~Makoff {\em et al.}, Phys. Rev. Lett. {\bf 70} 1591
        (1993). 
\bibitem{ber67} F.~A.~Berends, A.~Donnachie, and G.~C.~Oades, Phys. Lett. 
	{\bf B26} 109 (1967);~F.~A.~Berends, A.~Donnachie, and G.~C.~Oades, Phys. Rev. 
	{\bf 171} 1457 (1968).  
\bibitem{bol86} V.N.~Bolotov {\em et al.}, Sov. J. Nucl. Phys. {\bf 44} 68
        (1986). 
\bibitem{bar88} V.V.~Barmin{ \em et al.}, Sov. J. Nucl. Phys. {\bf 48} 1032
        (1988). 
\bibitem{main:99} M.~Abe {\em et al.}, Phys. Rev. Lett. {\bf 83} 4253 (1999).
\bibitem{nim} J.A.~Macdonald {\em et al.}, Nucl. Instr. and Meth. {\bf A506}
	60 (2003).
\bibitem{shi00} S.~Shimizu {\em et al.}, Phys. Lett. {\bf B495} 33
	(2000);~A.S.~Levchenko {\em et al.}, hep-ex/0111048;~Y.-H.~Shin 
	{\em et al.}, Eur. Phys. J. {\bf C12} 627 (2000);~K.~Horie 
	{\em et al.}, Phys. Lett. {\bf B513} 311 (2001).
\bibitem{ali03} ~M.A.~Aliev {\em et al.}, Phys. Lett. {\bf B554} 7 (2003).
\bibitem{amo99} G.~Amoros and J.~Bijnens, J. Phys. G {\bf 25} 1607
	(1999).
\bibitem{che60} G.F.~Chew and S.~Mandelstam, Phys. Rev. {\bf 119} 467 
	(1960).
\bibitem{daf92} G.~D'Ambrosio, M.~Miragliuolo, and P.~Santorelli, in
        $Da\Phi ne Physics Handbook$, edited by L.~Maiani, G.~Pancheri,
        and N.~Paver (Laboratori Nazionali di Frascati, Frascati, 1992).
\bibitem{pdb02} Particle Data Group, Phys. Rev. {\bf D66} 010001 (2002).
\end{references}
\end{document}